# Precisely positioned generation of $CsPbBr_3$ nano-light sources in a $Cs_4PbBr_6$ film by electron beam irradiation

*Tomoyasu Fujimaru[1], Kanta Hirai[2], Masato Inamata[1], Hiromu Tanaka[1], Midori Ikeuchi[3],*

*Hidehiro Yamashita[3], Mitsutaka Haruta[4], Takehiko Tamaoka[5], Naohiko Kawasaki[5],*

*and Hikaru Saito*[\*,3]*

[1] Interdisciplinary Graduate School of Engineering Sciences, Kyushu University, 6-1 Kasugakoen, Kasuga, Fukuoka 816-8580, Japan
[2] Department of Interdisciplinary Engineering, School of Engineering, Kyushu University, 6-1 Kasugakoen, Kasuga, Fukuoka 816-8580, Japan
[3] Institute for Materials Chemistry and Engineering, Kyushu University, 6-1 Kasugakoen, Kasuga, Fukuoka 816-8580, Japan
[4] Institute for Chemical Research, Kyoto University, Uji, Kyoto 611-0011, Japan
[5] Morphological Research Laboratory, Toray Research Center Inc., Otsu, Shiga 520-8567, Japan

*corresponding authors

**ABSTRACT**

Integration of high-quality photon emitters at specific locations within nanophotonic structures or optoelectronic devices is a key to innovating on-chip optical control and quantum technologies. Halide perovskite nanoparticles have great potential as single photon emitters with high quantum efficiency. To achieve their full potential, they must be embedded in a host material that ensures chemical stability and passivates surface defects. A previous experiment on a $CsPbBr_3$–$Cs_4PbBr_6$ nanocomposite film suggested possibility that electron beam irradiation can be used to control positions of $CsPbBr_3$ nano-light sources in the $Cs_4PbBr_6$ host although the effects of electron beam irradiation are not fully understood. Here, we fabricate a $Cs_4PbBr_6$–$CsBr$ film, not containing the $CsPbBr_3$ phase, and provide direct evidence that $CsPbBr_3$ nanoparticles can be locally generated in the $Cs_4PbBr_6$ host by irradiation with a focused electron beam. We further demonstrate perovskite nano-light source arrays with submicron spacing using this method.

**Table of Contents**

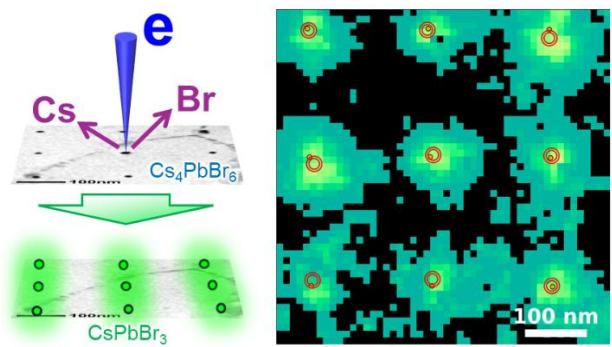

Precise positioning of photon emitters at specific locations within nanophotonic structures is essential for advanced light control or quantum technologies, such as strong light-matter interactions or Fano interferences,[1–3] and spin-selective photon routing.[4–6] Various top-down processes using electron beam lithography have been proposed for deterministic integration of quantum dots into on-chip optical circuits.[7–10] Trapping excitons by locally introduced strain into transition metal dichalcogenides using nanostructures is a promising strategy to arrange quantum emitters and plasmonic resonators in proximity, achieving single-photon emitter arrays with a highly enhanced emission rate[11] and a control of spin angular momentum of single photon emitters.[12] Hexagonal boron nitride is also attracting great attention as a robust and versatile host for single photon emitters which can be introduced at precise locations by nanoindentation,[13] irradiation with ion beams,[14] electron beams,[15] or lasers.[16]

However, such precise positioning of halide perovskite particles has not yet been reported, despite their excellent potential for highly efficient light-emitting diodes (LEDs),[17,18] lasers,[19,20] and single photon emitters functioning even at room temperature.[21,22] To achieve this, we must find a nanoscale process to directly generate halide perovskite nanoparticles in a host material that ensures their chemical stability and passivates surface defects causing performance degradation.[17] Tips to establish such a nanoscale process is seen in a previous experiment on a $CsPbBr_3$–$Cs_4PbBr_6$ nanocomposite film, where electron beam irradiation increased green light emission from $CsPbBr_3$ nanoparticles in the film.[23] If this is due to the generation of $CsPbBr_3$ nanoparticles in the $Cs_4PbBr_6$ grains, electron beam irradiation can be used to control the positions of $CsPbBr_3$ nano-light sources. $Cs_4PbBr_6$ having a relatively wide bandgap of ~3.9 eV, is an excellent host material for confining charge carriers within embedded $CsPbBr_3$ nanoparticles having a narrower bandgap of ~2.3 eV.[24] In fact, photoluminescence yields of over 90% were measured from $CsPbBr_3$ nanoparticles embedded in the $Cs_4PbBr_6$ host.[25] Furthermore, electroluminescence of $CsPbBr_3$–$Cs_4PbBr_6$ nanocomposites demonstrated in previous studies[26,27] also provides motivation to position-controlled $CsPbBr_3$ nano-light source generation in the $Cs_4PbBr_6$ host.

Here, we fabricate a film mainly composed of $Cs_4PbBr_6$ grains, excluding the $CsPbBr_3$ phase, and verify if $CsPbBr_3$ nanoparticles can be generated by irradiating $Cs_4PbBr_6$ with a focused electron beam. The changes in element ratios of the film caused by electron beam irradiation are revealed by energy-dispersive X-ray spectroscopy (EDS), the resulting nanoparticles are identified by electron energy-loss spectroscopy (EELS), and the light emission of the generated nanoparticles are characterized by cathodoluminescence (CL) spectroscopy.

## RESULTS AND DISCUSSION

### Characterization of fabricated films

In the previous experiment, green nano-light sources were formed by electron beam irradiation of $CsPbBr_3$–$Cs_4PbBr_6$ nanocomposite film.[23] One possible interpretation was that new $CsPbBr_3$ nanoparticles were generated in the $Cs_4PbBr_6$ grains. However, that experiment also provided another interpretation that the light emission efficiency of the initially existing $CsPbBr_3$ nanoparticles was improved by some effects of electron beam irradiation. To verify the former possibility, we must prepare sufficiently large $Cs_4PbBr_6$ grains not containing $CsPbBr_3$ nanoparticles inside them. In this study, we fabricated two types of halide films by thermal evaporation. One is composed of well-separated $Cs_4PbBr_6$ and $CsPbBr_3$ grains obtained by annealing the $CsPbBr_3$–$Cs_4PbBr_6$ nanocomposite film[23,28] in vacuum at 250 °C for 30 min. The other is composed of $Cs_4PbBr_6$ and $CsBr$ grains and free from the $CsPbBr_3$ phase. The details of film fabrication are provided in Method section. CL and EELS spectra obtained from $Cs_4PbBr_6$ and CsPbBr$_3$ grains in the former film were used as references when analyzing the changes of the latter film due to electron beam irradiation. The following scanning transmission electron microscopy (STEM)-based analysis was performed at a relatively low dose rate (0.4 e/nm$^2$/ns or less) and total dose (2 × 10$^7$ e/nm$^2$ or less), and no light source generation was recognized during this analysis.

Figure 1a shows a bright-field (BF) STEM image of the CsPbBr$_3$–Cs$_4$PbBr$_6$ film. The white areas are pores without the CsPbBr$_3$–Cs$_4$PbBr$_6$ layer. Such pores were not observed in the film without the post-annealing[23,28] and were therefore formed during the annealing. Figure 1b shows a CL map obtained from the same field of view using a band-pass filter with the wavelength range of 500–550 nm, revealing that this film contains grains of several hundred nanometers that emit light in this wavelength range. The CL spectrum obtained from one of these bright grains (green arrow in Figs. 1a and b) has a peak wavelength of 526 nm (Fig. 1c), which is close to that of previously measured from micrometer-scale $CsPbBr_3$ particles.[29] On the other hand, no significant spectral features were observed in the visible range from the surroundings, which is consistent with the film being composed primarily of nonradiative $Cs_4PbBr_6$ grains.[30] In the film obtained without post-annealing, the $CsPbBr_3$ nanoparticles were dispersed evenly throughout the film, and CL maps showed a uniform intensity distribution at the nanoscale.[23] The post-annealing introduced in this study resulted in a structure separated into $CsPbBr_3$ and $Cs_4PbBr_6$ grains with sizes from several hundred nanometers to micrometers, allowing estimation of the phase volume ratio of $CsPbBr_3$ to be 23% from the bright area ratio in the CL map (Fig. 1b).

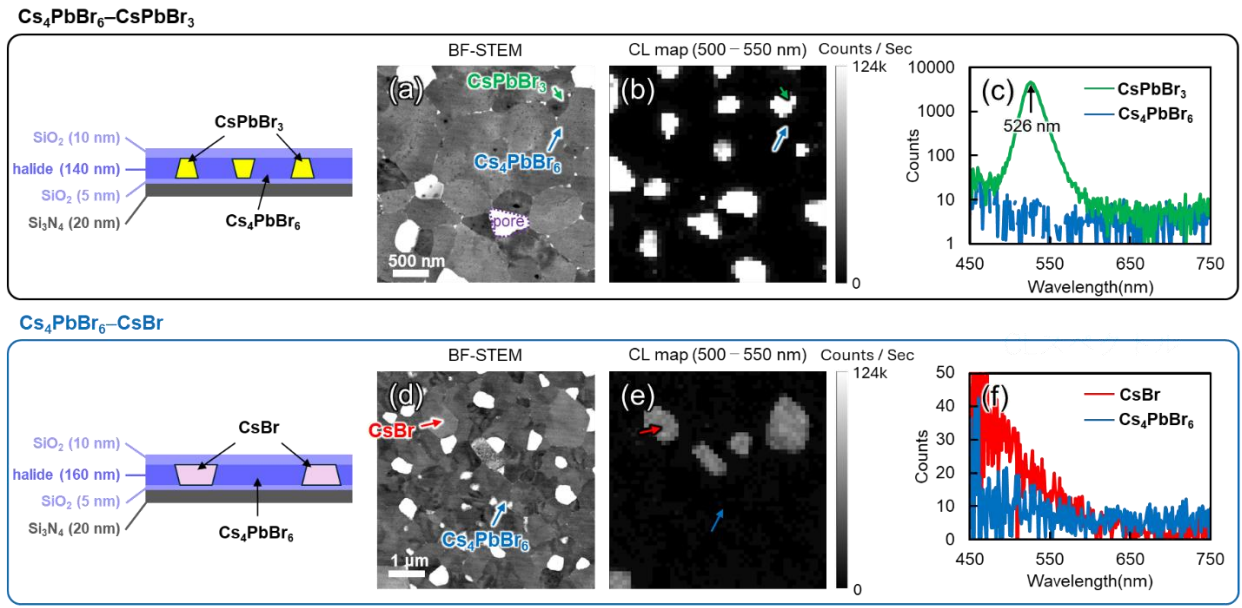

**Figure 1** Structural analysis of the fabricated CsPbBr$_3$–Cs$_4$PbBr$_6$ and Cs$_4$PbBr$_6$–CsBr films by CL spectroscopy. (a) BF-STEM image, (b) CL map, and (c) CL spectra of the CsPbBr$_3$–Cs$_4$PbBr$_6$ film. (d) BF-STEM image, (e) CL map, and (f) CL spectra of the Cs$_4$PbBr$_6$–CsBr film. The CL maps were obtained with wavelength filtering from 500 to 550 nm. The CL spectra were obtained while irradiating the arrowed positions. Schematic structures of the CsPbBr$_3$–Cs$_4$PbBr$_6$ and Cs$_4$PbBr$_6$–CsBr films are shown in insets on the left of (a) and (d), respectively. All the measurements here were performed at a beam energy of 300 keV and a probe current of 200 pA.

From the estimated $CsPbBr_3$ volume ratio, the amount of CsBr powder required to modify the element ratios of the evaporation source ($CsPbBr_3$–$Cs_4PbBr_6$ nanocomposite powder) to Cs:Pb:Br = 4:1:6 was calculated, and the second film ($Cs_4PbBr_6$–CsBr) was fabricated using the mixed evaporation source composed of $CsPbBr_3$–$Cs_4PbBr_6$ nanocomposite powder and CsBr powder. Figures 1d and e show a BF-STEM image and a wavelength-filtered CL map of this modified film, respectively. The CL maps show several submicron- to micrometer-sized grains that are brighter than the surroundings (e.g. red arrow in Fig. 1d), but are much darker than $CsPbBr_3$ grains (Fig. 1b). The CL spectrum obtained from one of these slightly bright grains shows a broad peak in the shorter wavelength side (Fig. 1f), which is clearly different from that of $CsPbBr_3$ grains (Fig. 1c). To identify the chemicals consisting of this modified film, we analyzed it using EELS and EDS. In the annular dark-filed (ADF)-STEM image, the above "slightly bright grains" appear in slightly darker intensity than the surroundings (e.g. area B in Fig. 2a). An EELS spectrum obtained from the surroundings (area A) shows peaks at 4 eV and 5.5 eV (Fig. 2b), which are consistent with those detected from $Cs_4PbBr_6$ particles.[28,31] On the other hand, the slightly bright grain (area B) does not show any characteristic peaks below 6 eV, and the presence of a shoulder at 6.7 eV is consistent with that detected from CsBr.[32] Elemental mapping by EDS shows that Pb is not detected in the slightly bright grains but instead the Cs density is increased compared to the surrounding area (Figs. 2c–f). Accordingly, by adjusting the element ratios as described above, this modified film is mainly composed of Cs$_4$PbBr$_6$ and free from CsPbBr$_3$ although it contains CsBr as a second phase probably due to an estimation error of the element ratios of the evaporation source. This film not only makes it straightforward to analyze the changes by electron beam irradiation, but also has advantages in the applications of nano-light sources since it does not contain background light sources.

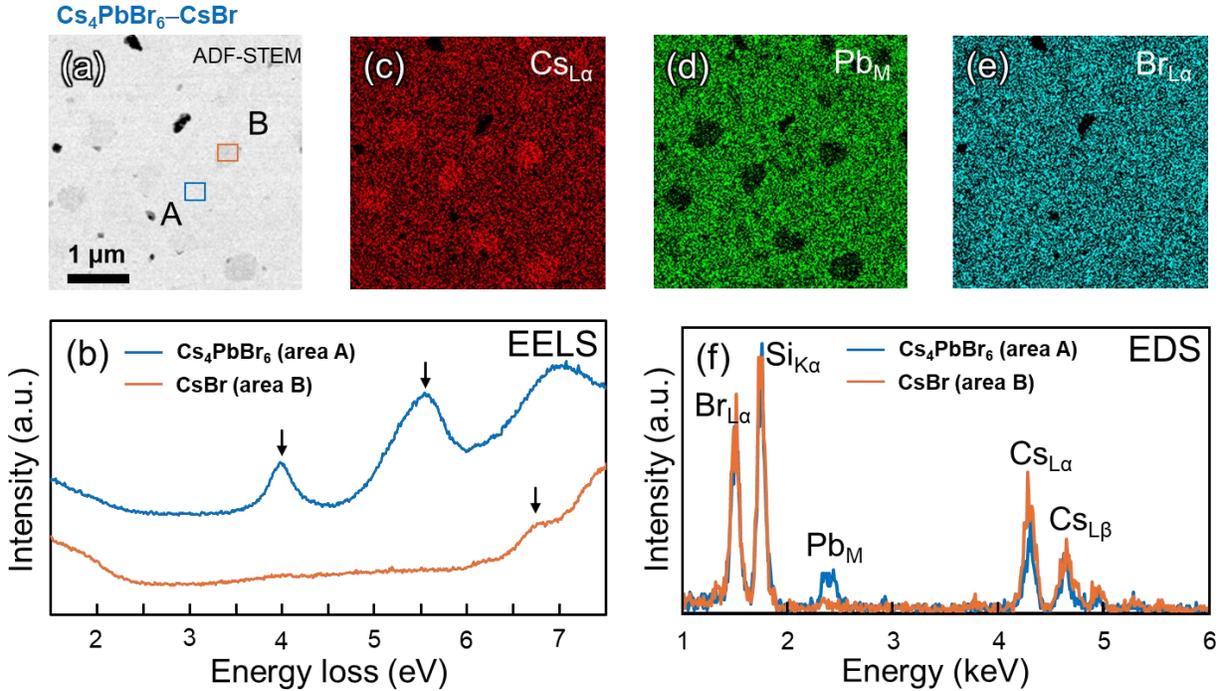

**Figure 2** Structural analysis of the Cs$_4$PbBr$_6$–CsBr film by EELS and EDS. (a) ADF-STEM image of the film. The blue (A) and orange (B) rectangles indicate extraction areas for the following EELS and EDS spectra. (b) EELS spectra obtained from a Cs$_4$PbBr$_6$ grain (area A) and a CsBr grain (area B). (c–e) Elemental maps using X-ray intensities at Cs$_{L\alpha}$, Pb$_M$, and Br$_{L\alpha}$ lines, respectively. (e) EDS spectra obtained from a Cs$_4$PbBr$_6$ grain (area A) and a CsBr grain (area B). All the measurements here were performed at a beam energy of 80 keV. The probe current was 18 pA for EELS and 50 pA for EDS.

**Generation of CsPbBr$_3$ nano-light sources in Cs$_4$PbBr$_6$ grains by electron beam irradiation**

Cs$_4$PbBr$_6$ grains in the prepared Cs$_4$PbBr$_6$–CsBr film were irradiated using a focused electron beam and the changes at the irradiation points were investigated by CL, EELS, and EDS. Intense irradiation to modify Cs$_4$PbBr$_6$ grains was performed at a beam energy of 300 keV under "high dose conditions" with a dose rate of more than 1.5 e/nm$^2$/ns and a total dose of more than 7.5 × 10$^9$ e/nm$^2$. STEM-based analysis was performed at a beam energy of 80 or 300 keV under "low dose conditions" with a dose rate of less than 0.4 e/nm$^2$/ns and a total dose of less than 2 × 10$^7$ e/nm$^2$.

Figures 3a shows the first CL map obtained from a Cs$_4$PbBr$_6$ grain region at a beam energy of 300 keV under a low dose condition with a beam current of 200 pA and a beam diameter of 2 nm, and no meaningful signal is detected in this region. After this mapping, the four points (green arrows) were irradiated using a stopped beam under a high dose condition with a beam current of

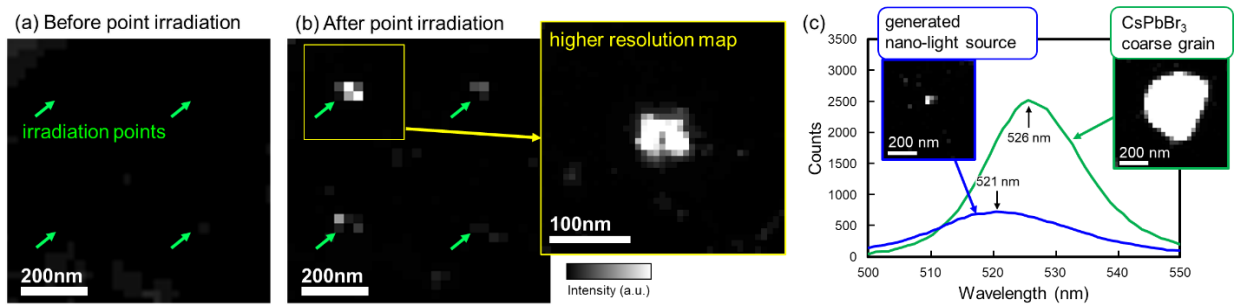

**Figure 3** Point irradiation to Cs$_4$PbBr$_6$ grains in the Cs$_4$PbBr$_6$–CsBr film with a focused electron beam and CL analysis of the generated nano-light sources. (a and b) CL maps (wavelength range: 500 to 550 nm) obtained before and after point irradiation. The green arrows indicate the four irradiation points. Each point was irradiated at a high dose rate of 1.5 e/nm$^2$/ns (beam current: 3 nA, beam diameter: 4 nm) for 15 seconds corresponding to a total dose of 2 × 10$^{10}$ e/nm$^2$. A higher resolution CL map was obtained around one irradiation point (right inset of b). (c) CL spectrum acquired during point irradiation for 10 seconds (blue curve) at the high dose rate, which is compared with that obtained from an initially existing CsPbBr$_3$ coarse grain in the CsPbBr$_3$–Cs$_4$PbBr$_6$ film (green curve). Wavelength-filtered CL maps of the generated nano-light source and the CsPbBr$_3$ coarse grain are shown in the insets. All the measurements here were performed at a beam energy of 300 keV and a probe current of 200 pA except for the CL spectrum acquired during the point irradiation (blue curve in c).

3 nA and a beam diameter of 4 nm for 15 seconds per point (s/point). The second CL map, shown in Fig. 3b, was then acquired from the same region under the same condition as the first map. Although the brightness and distribution vary point by point, bright spots appear at the four irradiation points while no significant changes are observed in the surrounding area. As discussed in the later subsection "Optimization of the irradiation time", an irradiation time on the order of seconds is required to generate nano-light sources under the high dose conditions used here. Figure 3c shows the CL spectrum obtained by collecting light emission during point irradiation of another $Cs_4PbBr_6$ grain under a high dose condition for 10 seconds (blue curve). The inset CL map was obtained under a low dose condition after the 10 seconds of point irradiation, showing a nano-light source generated and grown during the 10 seconds. The CL peak wavelength was 521 nm, which is close to that of an initially existing $CsPbBr_3$ coarse grain in the $CsPbBr_3$–$Cs_4PbBr_6$ film (green curve), but is slightly shifted to the shorter wavelength side due to the quantum size effect.[33,34] A similar blue shift has been observed in the CL and photoluminescence of $Cs_4PbBr_6$ containing $CsPbBr_3$ nanoparticles.[35,36]

We confirmed that the nano-light sources generated by electron beam irradiation are $CsPbBr_3$ nanoparticles by EELS measurements as discussed below. Figures 4a and b show a BF-STEM image and the corresponding wavelength-filtered CL map after irradiating nine points (green arrows) in $Cs_4PbBr_6$ grains under a high dose condition with a beam current of 8 nA and a beam diameter of 5 nm for 15 s/point. EELS mapping data was obtained from this area under a low dose condition, and two spectra were extracted near the central irradiation point (EELS1) and a surrounding area (EELS2) as shown in Fig. 4c. The spectra were analyzed using non-negative multiple linear least squares (MLLS) fitting assuming two components $CsPbBr_3$ and $Cs_4PbBr_6$. Each reference spectrum was obtained from the $CsPbBr_3$–$Cs_4PbBr_6$ film. This MLLS fitting was performed in the energy range of 1.4 to 6 eV, and in this energy range, no significant signals were expected from the $Si_3N_4$ or $SiO_2$ layers above and below the halide layer.[37,38] The MLLS fitting indicates that the experimental spectrum at the irradiation point (EELS1) is well reproduced by summing the reference spectra of $CsPbBr_3$ and $Cs_4PbBr_6$, confirming the generation of a $CsPbBr_3$ particle, while the spectrum obtained far from the irradiation point (EELS2) is almost reproduced only using the reference spectrum of $Cs_4PbBr_6$. The generated $CsPbBr_3$ nanoparticle is visualized as a CsPbBr$_3$ ratio map derived by the MLLS fitting at each measurement point, indicating the particle size of ~25 nm (Fig. 4d).

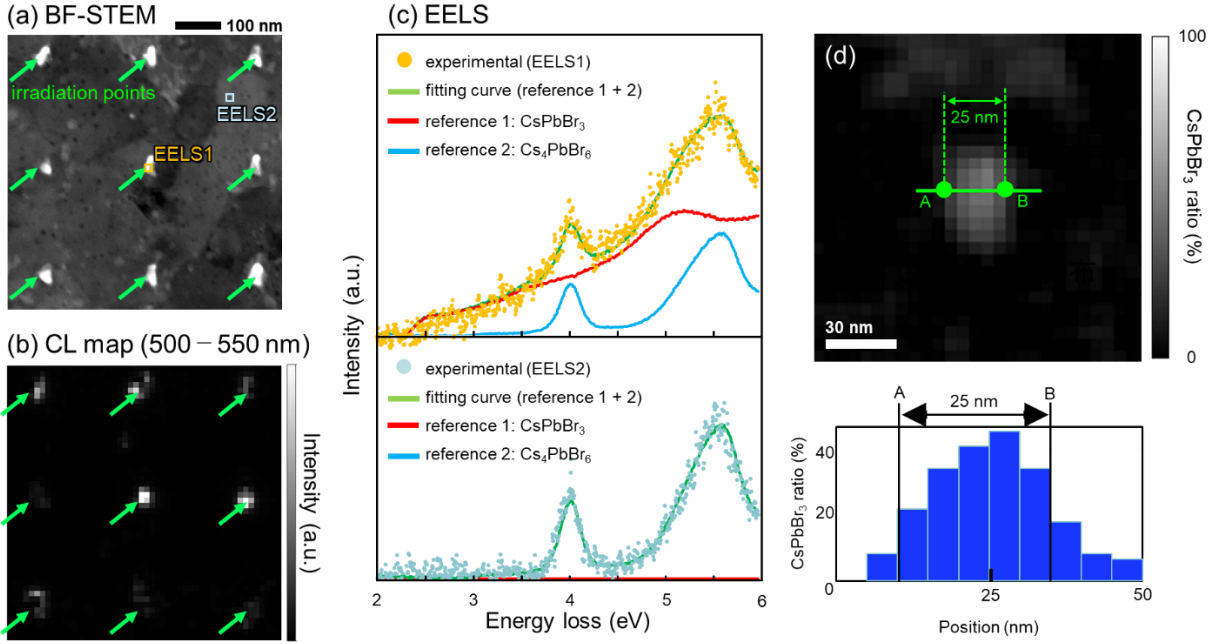

**Figure 4** EELS analysis of the generated nano-light source in the $Cs_4PbBr_6$–CsBr film. (a and b) BF-STEM image and CL map (wavelength range: 500 to 550 nm) obtained after point irradiation. The green arrows indicate the nine irradiation points. Each point was irradiated at a beam energy of 300 keV and a high dose rate of 2.5 $e/nm^2$/ns (beam current: 8 nA, beam diameter: 5 nm) for 15 s/point corresponding to a total dose of $4 \times 10^{10}$ $e/nm^2$. (c) EELS spectra obtained from the irradiation point at the center of b (top panel) and the surroundings (bottom panel), which are compared to MLLS fitting curves composed of $CsPbBr_3$ and $Cs_4PbBr_6$ reference spectra. A tail of zero-loss peak was subtracted by fitting with a power law function. Measurement areas for the EELS spectra are indicated by squares in a. (d) $CsPbBr_3$ ratio map derived by MLLS fitting at each measurement point in an EELS mapping including the central nano-light source in b. The bottom panel shows a profile along the line indicated in the ratio map. The BF-STEM image and the CL map in a and b were obtained at a beam energy of 300 keV and a probe current of 200 pA, and the EELS mapping data for c and d were obtained at a beam energy of 80 keV and a probe current of 18 pA.

At the irradiation points, the sample thickness (or density) decreases as suggested by the bright spots observed in the BF-STEM image (Fig. 4a), and the generated CsPbBr$_3$ nanoparticles have different element ratios from the original Cs$_4$PbBr$_6$ phase. These facts raise the possibility of knock-on of specific elements by irradiating Cs$_4$PbBr$_6$ with an electron beam. Such local changes in the element ratios were confirmed by using EDS mapping under a low dose condition for a field of view shown in Fig. 5a, where 9 points were irradiated under a high dose condition with a beam current of 3 nA and a beam diameter of 1 nm for 15 s/point in advance. Elemental maps using the X-ray intensities at the $Cs_{L\alpha}$, $Pb_M$, and $Br_{L\alpha}$ lines are shown in Figs. 5b–d, respectively. Due to the decrease in X-ray intensity, irradiation points (white arrows) are clearly recognized as dark spots in the Cs map (Fig. 5b) and Br map (Fig. 5d), while the intensity decrease is minor for Pb (Fig. 5c). This difference is clearer in a comparison of the EDS spectra; the $Cs_{L\alpha}$ and $Br_{L\alpha}$ peaks in the averaged spectrum extracted from the nine irradiation points are lower than those in the reference spectrum obtained from $Cs_4PbBr_6$ grains far from this field of view (Fig. 5e). Quantitative comparison normalized by $Pb_M$ peak intensity reveals that the peak intensities of $Cs_{L\alpha}$ and $Br_{L\alpha}$ decrease by ~20% during the irradiation (Fig. 5f). This result indicates that electron beam irradiation preferentially knocks out Cs and Br atoms from $Cs_4PbBr_6$, locally shifting the element ratios to Cs:Pb:Br = 1:1:3, stabilizing the $CsPbBr_3$ phase. Note that excessive irradiation can damage $CsPbBr_3$ nanoparticles as seen in a previous study.[39]

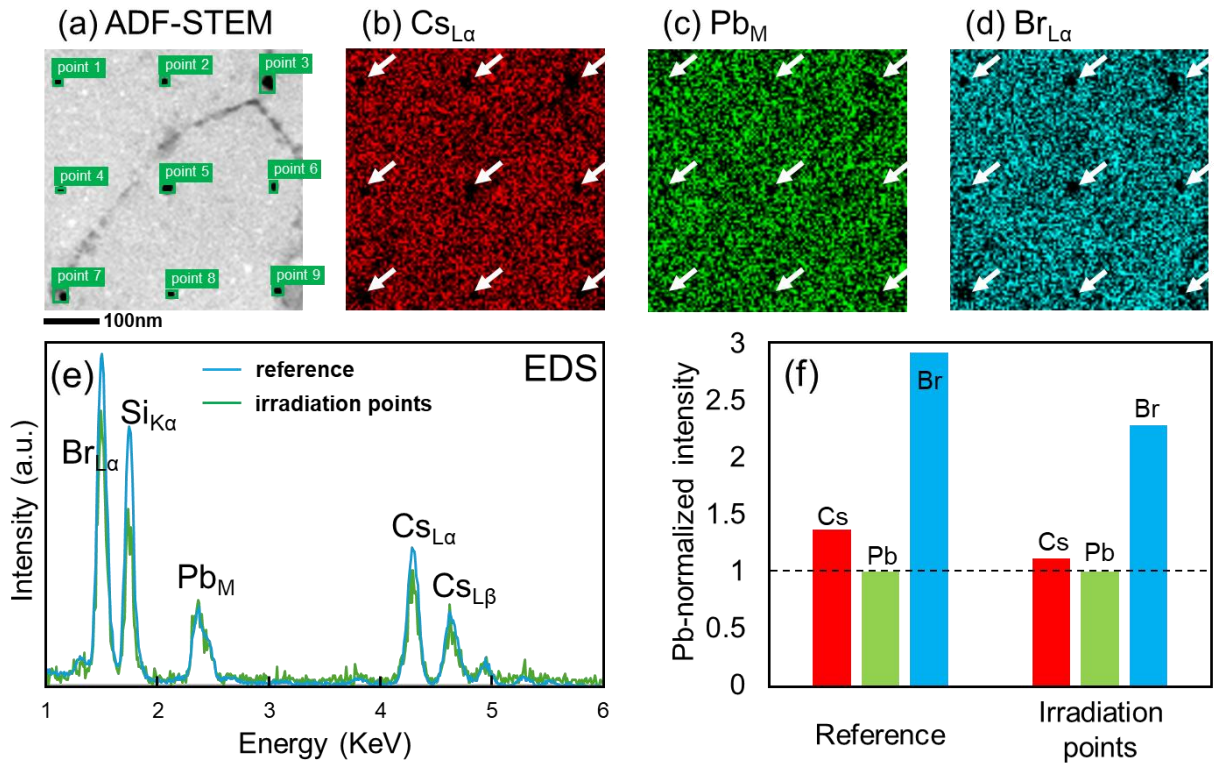

**Figure 5** EDS analysis of the local element ratio change due to electron beam irradiation in the Cs$_4$PbBr$_6$–CsBr film. (a) BF-STEM image obtained after point irradiation at a beam energy of 300 keV and a high dose rate of 6 e/nm$^2$/ns (beam current: 3 nA, beam diameter: 1 nm) for 15 s/point corresponding to a total dose of 9 × 10$^{10}$ e/nm$^2$. (b–d) Elemental maps using X-ray intensities at Cs$_{L\alpha}$, Pb$_M$, and Br$_{L\alpha}$ lines, respectively. The white arrows indicate the nine irradiation points. (e) EDS spectra averaged over nine extraction areas near the irradiation points (green curve) and extracted from reference Cs$_4$PbBr$_6$ grains far from this field of view (aqua curve). The nine extraction areas are indicated by rectangles in a. (f) Change in X-ray intensities between the two spectra in e, quantified by fitting Gaussian curves to Cs$_{L\alpha}$, Pb$_M$, and Br$_{L\alpha}$ peaks. The intensities are normalized by Pb$_M$ peak intensity. All the measurements here were performed at a beam energy of 300 keV and a probe current of 50 pA.

**Optimization of the irradiation time**

To optimize the irradiation time and evaluate the performance of our proposed method, we obtained CL maps after irradiation for different irradiation times as shown in Figs. 6a–e. Irradiation points were selected on a 3 × 3 square lattice with a lattice spacing of ~210 nm, and each point was irradiated under a high dose condition with a beam current of 3 nA and a beam diameter of 4 nm for 5–25 seconds. Wavelength-filtered CL maps acquired at a low dose condition current are shown in Figs. 6a–e. When the irradiation time is 5 s/point, clear bright spots are observed at less than half of the points (Fig. 6a). At 10 s/point or more, bright spots were formed at all the points, and the spots become brighter overall with increasing the irradiation time (Figs. 6b–d). However, some spots become significantly darker at 25 s/point (Fig. 6d) compared to 20 s/point, suggesting that the formed $CsPbBr_3$ nanoparticles are damaged by excessive irradiation.[39] To optimize the irradiation time, we evaluated the quality of each nano-light source generation as follows. The irradiation points were first estimated from BF-STEM images acquired after the irradiation. As mentioned above, the film thickness decreases due to the point irradiation, and the irradiation traces were observed as bright spots in the BF-STEM images (e.g. Fig. 4a). The positions of irradiation centers are determined from the irradiation traces with an image process provided in Supporting Information. The brightness of each generated nano-light source is evaluated by the total CL intensity within the unit cell (square of ~210 nm side length) centered on the irradiation center. The center of nano-light source $\bar{\mathbf{r}}$ is defined as the weighted average $\bar{\mathbf{r}} = \sum I_{i,j} \mathbf{r}_{i,j} / \sum I_{i,j}$ using CL intensity $I_{i,j}$ and position $\mathbf{r}_{i,j}$ at each pixel within the unit cell. The radius of nano-light source $\Delta r$ is defined as mean deviation $\Delta r = \sum I_{i,j} |\mathbf{r}_{i,j} - \bar{\mathbf{r}}| / \sum I_{i,j}$, which is calculated within the unit cell. We here evaluate the nano-light source generation as "successful" when the total CL intensity exceeds $10^2$, the light source center $\bar{\mathbf{r}}$ is located within 30 nm from the irradiation center, and the radius $\Delta r$ is less than 40 nm or less. Among these, particularly bright sources with a total CL count exceeding $10^3$ are evaluated as "very successful". These evaluations are indicated by marking a "○" and a "◎" at each center $\bar{\mathbf{r}}$ in the CL maps, respectively. When the above criteria are not met, an "×" is marked. According to the above evaluation criteria, an irradiation time of ~20 seconds corresponding to a total dose of ~$3 \times 10^{10}$ e/nm$^2$ is optimal.

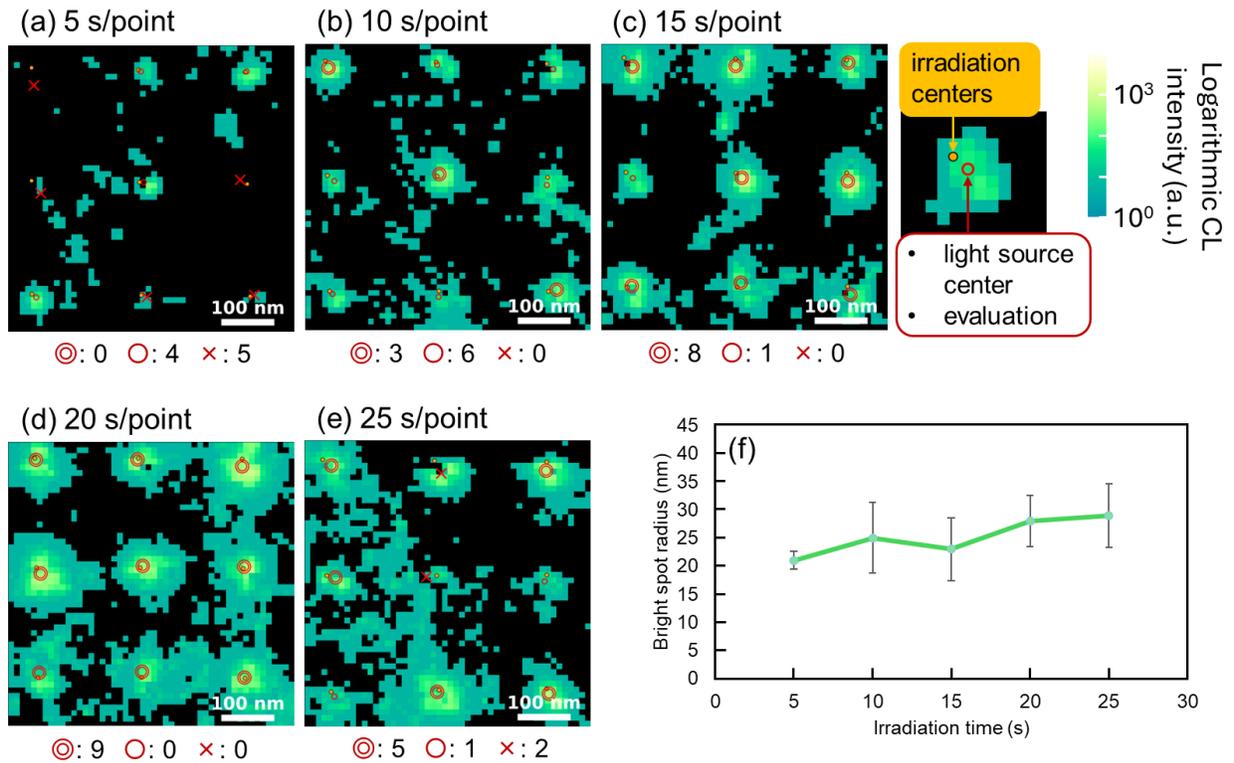

**Figure 6** Optimization of the irradiation time. (a–e) CL maps (wavelength range: 500 to 550 nm) obtained from $Cs_4PbBr_6$ grains in the $Cs_4PbBr_6$–$CsBr$ film after point irradiation. Irradiation points were selected on a 3 × 3 square lattice with a lattice spacing of ~210 nm. Each point was irradiated with a beam energy of 300 keV, a beam current of 3 nA, and a beam diameter of 4 nm for 5–25 seconds. The inset on the right of c describes the marks near each irradiation point. The orange dots indicate the irradiation centers measured at the irradiation traces appearing in BF-STEM images obtained after the point irradiation. Quality of each point irradiation is classified into 3 levels "◎" (very successful), "○" (successful), and "×" (failed) according to the criteria described in the main text, and these symbols are marked at each light source center $\bar{r}$ in the CL maps. The number assigned to each level is displayed below each CL map (f) Irradiation time dependence of average radius of generated nano-light sources. The failed points are excluded in this averaging. The error bars represent the standard deviation. All the measurements here were performed at a beam energy of 300 keV and a probe current of 200 pA.

      The average radius of nano-light sources increases gradually with irradiation time, ranging from 20 nm to 30 nm (Fig. 6f). The standard deviation between the irradiation center and the light source center was evaluated to be 10 nm for irradiation time of 20 s/point. Therefore, the proposed method enables positioning light sources with a lateral size of several tens of nanometers and the lateral positional precision of 10 nm. The increase in source radius with irradiation time suggests that the source radius can be made smaller by controlling the irradiation in a shorter time range or a lower dose rate range. However, achieving such finer light source positioning requires development of an irradiation control system tailored to individual nucleation, including feedback of real-time CL measurements, since the time at which nucleation begins is highly fluctuated as suggested by the CL map at irradiation time of 5 s/point (Fig. 6a).

## CONCLUSION

We fabricated the film mainly composed of $Cs_4PbBr_6$ excluding $CsPbBr_3$, and demonstrated that a $CsPbBr_3$ nano-light source can be generated by irradiating $Cs_4PbBr_6$ with an electron beam. Our elemental analysis suggested that the electron beam preferentially knocked out Cs and Br atoms from $Cs_4PbBr_6$, leading to the generation of $CsPbBr_3$ nanoparticles. Using this proposed method, we further demonstrated that nano-light sources with radii of several tens of nanometers can be arranged with the positional precision of 10 nm and submicron spacing. This method is based on evaporation and electron beam irradiation processes, and thus, is versatile enough to be integrated in various applications including on-chip optical devices.

## METHODS

### Fabrication of $CsPbBr_3$–$Cs_4PbBr_6$ and $Cs_4PbBr_6$–CsBr films

We first fabricated a $CsPbBr_3$–$Cs_4PbBr_6$ nanocomposite film on a $Si_3N_4$ (20 nm thick) film substrate (SiMPore, USA) in a similar manner as previous studies.[23,28] After depositing $SiO_2$ layer (5 nm thick) on the $Si_3N_4$ film substrate by radio frequency (RF) magnetron sputtering to obtain a clean surface, $CsPbBr_3$–$Cs_4PbBr_6$ nanocomposite layer (140 nm thick) was deposited by thermal evaporation. The evaporation source for the $CsPbBr_3$–$Cs_4PbBr_6$ deposition was synthesized based on the protocol proposed previously.[25] $PbBr_2$ (0.22 mmol, ≥ 98%, Sigma-Aldrich) and cesium acetate (0.88 mmol, 99.9% trace metals basis, Sigma-Aldrich) were stirred in dimethyl sulfoxide (0.5 ml, anhydrous, ≥ 99.9%, Sigma-Aldrich) for 1 h. After adding 0.1 mL aqueous HBr solution (48% by weight in $H_2O$, Sigma-Aldrich), the mixed solution was stirred for 12 h. The precipitation was separated from the mixed solution by centrifugation, and the powder was dried under vacuum overnight. To prevent degradation due to air exposure, the top surface of the $CsPbBr_3$–$Cs_4PbBr_6$ nanocomposite layer was covered with a 10 nm thick $SiO_2$ layer by RF magnetron sputtering. Overall, a $SiO_2$ (10 nm)/$CsPbBr_3$–$Cs_4PbBr_6$ (140 nm)/ $SiO_2$ (5 nm)/ $Si_3N_4$ (20 nm) multilayer film was fabricated. All the above procedures were done at room temperature. Then, this multilayer film was annealed in vacuum at 250 °C for 30 min to obtain a structure separated into $CsPbBr_3$ and $Cs_4PbBr_6$ grains with sizes from several hundred nanometers to micrometers as shown in Figs. 1a and b.

The $Cs_4PbBr_6$–$CsBr$ film used in this study was fabricated with the same process as above, except that the element ratios were tuned by adding 23 wt% $CsBr$ powder to the $CsPbBr_3$–$Cs_4PbBr_6$ evaporation source. Overall, a $SiO_2$ (10 nm)/ $Cs_4PbBr_6$–$CsBr$ (160 nm)/ $SiO_2$ (5 nm)/ $Si_3N_4$ (20 nm) multilayer film was fabricated. This multilayer film was also annealed in vacuum at 250 °C for 30 min.

**STEM-based spectroscopy and point irradiation with a focused electron beam**

CL analysis was performed at a beam energy of 300 keV, a beam current of 200 pA, and a beam diameter of 2 nm using a modified transmission electron microscope JEM-3200FSK (JEOL). The details of the CL optics are provided in a previous study.[28] A CL mapping system was newly implemented in this study, which is based on a single photon counting detector linked to the beam scan. The exposure time per pixel was 50 ms. A bandpass filter with a wavelength range of 500 to 550 nm was used to highlight $CsPbBr_3$ nanoparticles or grains.

EELS and EDS analyses were performed at beam energies of 80 or 300 keV, a beam current of 18 or 50 pA, and a sub-nanometer beam diameter using a transmission electron microscope Titan Cubed G2 60-300 (Thermo Fisher Scientific, USA) equipped with an aberration corrector for the probe forming lens system, a monochromator and an energy filter Quantum 965 (Gatan, USA) for EELS, and four-quadrant windowless super-X silicon drift detectors for EDS. By using the monochromator, the energy resolution was improved to 0.14 eV measured from the full width at half maximum of zero-loss peak (ZLP). The exposure time per pixel for EELS mapping was 50 ms. The typical dwell time for EDS mapping was on the order of 10 microseconds.

Point irradiation was mainly performed with a beam energy of 300 keV, beam currents of 3 or 8 nA, and beam diameters of 4 or 5 nm using JEM-3200FSK. Only for EDS analysis (Fig. 5), point irradiation was performed with a beam energy of 300 keV, a beam current of 3 nA, and a beam diameter of 1 nm using Titan Cubed G2 60-300.

All the experiments using electron microscopes were performed at room temperature.

## CONFLICT OF INTEREST

The authors declare no competing financial interest.

## ACKNOWLEDGMENTS

We would like to thank Keiichirou Akiba and Takumi Sannomiya for valuable discussion on this study. Annealing of the samples was supported by Shiro Ihara and Mitsuhiro Murayama. This study was supported by JSPS KAKENHI Grant Numbers JP25K01640, JP23K17350, JP23K23196, JP22H05034. We would also like to mention that preliminary experiments of this study were conducted at the Institute for Chemical Research, Kyoto University, supported by "Advanced Research Infrastructure for Materials and Nanotechnology in Japan (ARIM)" of the Ministry of Education, Culture, Sports, Science and Technology (MEXT) and International Collaborative Research Program of Institute for Chemical Research, Kyoto University (grant #

2025-141).

# REFERENCES

1. Santhosh, K.; Bitton, O.; Chuntonov, L.; Haran, G. Vacuum Rabi Splitting in a Plasmonic Cavity at the Single Quantum Emitter Limit. *Nat. Commun.* **2016**, *7*, 11823.

2. Imada, H.; Miwa, K.; Imai-Imada, M.; Kawahara, S.; Kimura, K.; Kim, Y. Single-Molecule Investigation of Energy Dynamics in a Coupled Plasmon-Exciton System. *Phys. Rev. Lett.* **2017**, 119, 013901.

3. Leng, H.; Szychowski, B.; Daniel, M.-C.; Pelton, M. Strong Coupling and Induced Transparency at Room Temperature with Single Quantum Dots and Gap Plasmons. *Nat. Commun.* **2018**, 9, 4012.

4. Söllner, I.; Mahmoodian, S.; Hansen, S. L.; Midolo, L.; Javadi, A.; Kiršanskė, G.; Pregnolato, T.; El-Ella, H.; Lee, E. H.; Song, J. D.; Stobbe, S.; Lodahl, P. Deterministic Photon–Emitter Coupling in Chiral Photonic Circuits. *Nat. Nanotechnol.* **2015**, 10, 775–778.

5. Barik, S.; Karasahin, A.; Flower, C.; Cai, T.; Miyake, H.; DeGottardi, W.; Hafezi, M.; Waks, E. A Topological Quantum Optics Interface. *Science* **2018**, 359, 666–668.

6. Mehrabad, M. J.; Foster, A. P.; Dost, R.; Clarke, E.; Patil, P. K.; Fox, A. M.; Skolnick, M. S.; Wilson, L. R. Chiral Topological Photonics with an Embedded Quantum Emitter. *Optica* **2020**, 7, 1690–1696.

7. Esmaeil Zadeh, I.; Elshaari, A. W.; Jöns, K. D.; Fognini, A.; Dalacu, D.; Poole, P. J.; Reimer, M. E.; Zwiller, V. Deterministic Integration of Single Photon Sources in Silicon Based Photonic Circuits. *Nano Lett.* **2016**, 16, 2289–2294.

8. Davanco, M.; Liu, J.; Sapienza, L.; Zhang, C.-Z.; De Miranda Cardoso, J. V.; Verma, V.; Mirin, R.; Nam, S. W.; Liu, L.; Srinivasan, K. Heterogeneous Integration for On-Chip Quantum Photonic Circuits with Single Quantum Dot Devices. *Nat. Commun.* **2017**, 8, 889.

9. Schnauber, P.; Schall, J.; Bounouar, S.; Höhne, T.; Park, S.-I.; Ryu, G.-H.; Heindel, T.; Burger, S.; Song, J.-D.; Rodt, S.; Reitzenstein, S. Deterministic Integration of Quantum Dots into on-Chip Multimode Interference Beamsplitters Using in Situ Electron Beam Lithography. *Nano Lett.* **2018**, 18, 2336–2342.

10. Li, S.; Yang, Y.; Schall, J.; von Helversen, M.; Palekar, C.; Liu, H.; Roche, L.; Rodt, S.; Ni, H.; Zhang, Y.; Niu, Z.; Reitzenstein, S. Scalable Deterministic Integration of Two Quantum Dots into an On-Chip Quantum Circuit. *ACS Photonics* **2023**, 10, 2846–2853.

11. Luo, Y.; Shepard, G. D.; Ardelean, J. V.; Rhodes, D. A.; Kim, B.; Barmak, K.; Hone, J. C.; Strauf, S. Deterministic Coupling of Site-Controlled Quantum Emitters in Monolayer WSe$_2$ to Plasmonic Nanocavities. *Nat. Nanotechnol.* **2018**, 13, 1137–1142.

12. Lee, S.-J.; So, J.-P.; Kim, R. M.; Kim, K.-H.; Rha, H.-H.; Na, G.; Han, J. H.; Jeong, K.-Y.; Nam, K. T.; Park, H.-G. Spin Angular Momentum–Encoded Single-Photon Emitters in a Chiral Nanoparticle-Coupled WSe$_2$ Monolayer. *Sci. Adv.* **2024**, 10, eadn7210.

13. Luo, M.; Ge, J.; Huang, P.; Yu, Y.; Seo, I. C.; Lu, K.; Sun, H.; Tan, J. K.; Tay, B. K.; Kim, S.; Gao, W.; Li, H.; Nam, D. Deterministic Formation of Carbon-Functionalized Quantum Emitters in Hexagonal Boron Nitride. *Nat. Commun.* **2025**, 16, 11450.

14. Glushkov, E.; Macha, M.; Räth, E.; Navikas, V.; Ronceray, N.; Cheon, C. Y.; Ahmed, A.; Avsar, A.; Watanabe, K.; Taniguchi, T.; Shorubalko, I.; Kis, A.; Fantner, G.; Radenovic, A. Engineering Optically Active Defects in Hexagonal Boron Nitride Using Focused Ion Beam and Water. *ACS Nano* **2022**, 16, 3695–3703.

15. Gale, A.; Li, C.; Chen, Y.; Watanabe, K.; Taniguchi, T.; Aharonovich, I.; Toth, M. Site-Specific Fabrication of Blue Quantum Emitters in Hexagonal Boron Nitride. *ACS Photonics* **2022**, 9, 2170–2177.

16. Gan, L.; Zhang, D.; Zhang, R.; Zhang, Q.; Sun, H.; Li, Y.; Ning, C.-Z. Large-Scale, High-Yield Laser Fabrication of Bright and Pure Single-Photon Emitters at Room Temperature in Hexagonal Boron Nitride. *ACS Nano* **2022**, 16, 14254–14261.

17. Lin, K.; Xing, J.; Quan, L. N.; García de Arquer, F. P.; Gong, X.; Lu, J.; Xie, L.; Zhao, W.; Zhang, D.; Yan, C.; Li, W.; Liu, X.; Lu, Y.; Kirman, J.; Sargent, E. H.; Xiong, Q.; Wei, Z. Perovskite light-emitting diodes with external quantum efficiency exceeding 20 per cent. *Nature* **2018**, 562, 245–248.

18. Lian, Y.; Wang, Y.; Yuan, Y.; Ren, Z.; Tang, W.; Liu, Z.; Xing, S.; Ji, K.; Yuan, B.; Yang, Y.; Gao, Y.; Zhang, S.; Zhou, K.; Zhang, G.; Stranks, S. D.; Zhao, B.; Di, D. Downscaling micro- and nano-perovskite LEDs. *Nature* **2025**, 640, 62–68.

19. Schlaus, A. P.; Spencer, M. S.; Miyata, K.; Liu, F.; Wang, X.; Datta, I.; Lipson, M.; Pan, A.; Zhu, X.-Y. How lasing happens in CsPbBr$_3$ perovskite nanowires. *Nat. Commun.* **2019**, 10, 265.

20. Tiguntseva, E.; Koshelev, K.; Furasova, A.; Tonkaev, P.; Mikhailovskii, V.; Ushakova, E. V.; Baranov, D. G.; Shegai, T.; Zakhidov, A. A.; Kivshar, Y.; Makarov, S. V. Room-Temperature Lasing from Mie-Resonant Nonplasmonic Nanoparticles. *ACS Nano* **2020**, 14, 8149–8156.

21. Park, Y.-S.; Guo, S.; Makarov, N. S.; Klimov, V. I. Room-Temperature Single-Photon Emission from Individual Perovskite Quantum Dots. *ACS Nano* **2015**, 9, 10386–10393.

22. Murali, R.; Panda, M. K.; Challa, R. K.; Acharjee, D.; Rao Soma, V.; Ghosh, S.; Raavi, S. S. K. Bright and Stable Single-Photon Emission in Zinc-Alloyed CsPbBr$_3$ Nanocrystals Through Controlled Auger Recombination. *Small* **2025**, 22, e05011.

23. Saito, H.; Kihara, K.; Ikeuchi, M.; Yanagimoto, S.; Kubota, T.; Akiba, K.; Sannomiya, T. Nano-light source generation by electron beam irradiation of CsPbBr$_3$/Cs$_4$PbBr$_6$ composites. *Appl. Phys. Lett.* **2025**, 127, 261102.

24. Kang, B.; Biswas, K. Exploring Polaronic, Excitonic Structures and Luminescence in Cs$_4$PbBr$_6$/CsPbBr$_3$. *J. Phys. Chem. Lett.* **2018**, 9, 830–836.

25. Chen, Y.-M.; Zhou, Y.; Zhao, Q.; Zhang, J.-Y.; Ma, J.-P.; Xuan, T.-T.; Guo, S.-Q.; Yong, Z.-J.; Wang, J.; Kuroiwa, Y.; Moriyoshi, C.; Sun, H.-T. *Cs$_4$PbBr$_6$/CsPbBr$_3$ Perovskite Composites with Near-Unity Luminescence Quantum Yield: Large-Scale Synthesis, Luminescence and Formation Mechanism, and White Light-Emitting Diode Application.* *ACS Appl. Mater. Interfaces* **2018**, 10, 15905–15912.

26. Xu, J.; Huang, W.; Li, P.; Onken, D. R.; Dun, C.; Guo, Y.; Ucer, K. B.; Lu, C.; Wang, H.; Geyer, S. M.; Williams, R. T.; Carroll, D. L. Imbedded Nanocrystals of CsPbBr$_3$ in Cs$_4$PbBr$_6$: Kinetics, Enhanced Oscillator Strength, and Application in Light-Emitting Diodes. *Adv. Mater.* **2017**, 29, 1703703.

27. Zheng, S.; Wang, Z.; Jiang, N.; Huang, H.; Wu, X.; Li, D.; Teng, Q.; Li, J.; Li, C.; Li, J.; Pang, T.; Zeng, L.; Zhang, R.; Huang, F.; Lei, L.; Wu, T.; Yuan, F.; Chen, D. Ultralow voltage–driven efficient and stable perovskite light-emitting diodes. *Sci. Adv.* **2024**, 10, eadp8473.

28. Fujimaru, T.; Tanaka, H.; Inamata, M.; Ikeuchi, M.; Yamashita, H.; Miyazaki, H.; Gondo, T.; Hata, S.; Murayama, M.; Saito, H. Light Emission Enhancement on Nanostructured

Surfaces Quantitatively Evaluated by Cathodoluminescence Coincidence Counting. *ACS Photonics* **2025**, 12, 3073–3081.

29. Nekita, S.; Yanagimoto, S.; Sannomiya, T.; Akiba, K.; Takiguchi, M.; Sumikura, H.; Takagi, I.; Nakamura, K. G.; Yip, S. P.; Meng, Y.; Ho, J. C.; Okuyama, T.; Murayama, M.; Saito, H. Diffusion-Dominated Luminescence Dynamics of $CsPbBr_3$ Studied Using Cathodoluminescence and Microphotoluminescence Spectroscopy. *Nano Lett.* **2024**, 24, 3971–3977.

30. Riesen, N.; Lockrey, M.; Badek, K.; Riesen, H. On the origins of the green luminescence in the "zero-dimensional perovskite" $Cs_4PbBr_6$: conclusive results from cathodoluminescence imaging. *Nanoscale* **2019**, 11, 3925.

31. de Weerd, C.; Lin, J.; Gomez, L.; Fujiwara, Y.; Suenaga, K.; Gregorkiewicz, T. Hybridization of Single Nanocrystals of $Cs_4PbBr_6$ and $CsPbBr_3$. *J. Phys. Chem. C* **2017**, 121, 19490–19496.

32. Kiguchi, M.; Entani, S.; Saiki, K.; Koma, A. Atomic and electronic structure of CsBr film grown on LiF and KBr. *Surf. Sci.* **2003**, 523, 73–79.

33. Xu, F.; Kong, X.; Wang, W.; Juan, F.; Wang, M.; Wei, H.; Li, J.; Cao, B. Quantum size effect and surface defect passivation in size-controlled $CsPbBr_3$ quantum dots. *J. Alloys Compd.* **2020**, 831, 154834.

34. Chen, J.; Žídek, K.; Chábera, P.; Liu, D.; Cheng, P.; Nuuttila, L.; Al-Marri, M. J.; Lehtivuori, H.; Messing, M. E.; Han, K.; Zheng, K.; Pullerits, T. Size- and wavelength-dependent two-photon absorption cross-section of $CsPbBr_3$ perovskite quantum dots. *J. Phys. Chem. Lett.* **2017**, 8, 2316–2321.

35. Kubota, T.; Yanagimoto, S.; Saito, H.; Akiba, K.; Ishii, A.; Sannomiya, T. Cathodoluminescence spectral and lifetime mapping of $Cs_4PbBr_6$: fast lifetime and its scintillator application. *Appl. Phys. Express* **2024**, 17, 015005.

36. Cao, F.; Yu, D.; Ma, W.; Xu, X.; Cai, B.; Yang, Y. M.; Liu, S.; He, L.; Ke, Y.; Lan, S.; Choy, K.-L.; Zeng, H. Shining Emitter in a Stable Host: Design of Halide Perovskite Scintillators for X-ray Imaging from Commercial Concept. *ACS Nano* **2020**, 14, 5183–5193.

37. Asano, T.; Tezura, M.; Saitoh, M.; Tanaka, H.; Kikkawa, J.; Kimoto, K. Nanoscale observation of subgap excitations in β-Si3N4 with a high refractive index using low-voltage monochromated STEM: a new approach to analyze the physical properties of defects in dielectric materials. *Appl. Phys. Express* **2022**, 15, 076501.

38. Park, J.; Heo, S.; Chung, J.-G.; Kim, H.; Lee, H.; Kim, K.; Park, G.-S. Bandgap measurement of thin dielectric films using monochromated STEM-EELS. *Ultramicroscopy* **2009**, 109, 1183–1188.

39. Dang, Z.; Shamsi, J.; Palazon, F.; Imran, M.; Akkerman, Q. A.; Park, S.; Bertoni, G.; Prato, M.; Brescia, R.; Manna, L. In Situ Transmission Electron Microscopy Study of Electron Beam-Induced Transformations in Colloidal Cesium Lead Halide Perovskite Nanocrystals. *ACS Nano* **2017**, 11, 2124–2132.

# SUPPORTING INFORMATION

A.  **Positions of irradiation centers**

The positions of irradiation centers shown in Figs. 6a–e of the main text were estimated from irradiation traces observed in BF-STEM images. The procedure is explained below using an example of the irradiation time of 20 s/points. Figure S1a shows a BF-STEM image acquired after the point irradiation, in which nine irradiation traces are arranged in the form of a 3 × 3 square lattice. These illumination traces were clearly separated from the background by binarization (Fig. S1b). The positions of irradiation centers at the four corners (circled in yellow) were estimated from the average position calculated within each bright spot, and their midpoints were assumed to be the positions of the other five irradiation centers.

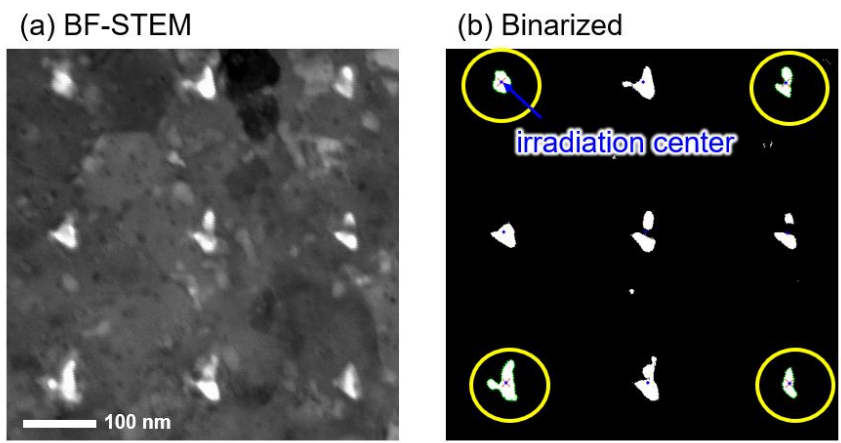

**Figure S1** (a) BF-STEM image acquired after point irradiation for an irradiation time of 20 s/points. (b) Binarized image of the BF-STEM image in (a). The estimated positions of irradiation centers are indicated by blue dots.